%% file: main.tex
\documentclass[sigconf, screen, nonacm]{acmart} 

\AtBeginDocument{%
  \providecommand\BibTeX{{%
    \normalfont B\kern-0.5em{\scshape i\kern-0.25em b}\kern-0.8em\TeX}}}

\copyrightyear{2021} 
\acmYear{2021} 
\setcopyright{acmcopyright}\acmConference[CHI '21]{CHI Conference on Human Factors in Computing Systems}{May 8--13, 2021}{Yokohama, Japan}
\acmBooktitle{CHI Conference on Human Factors in Computing Systems (CHI '21), May 8--13, 2021, Yokohama, Japan}
\acmPrice{15.00}
\acmDOI{10.1145/3411764.3445689}
\acmISBN{978-1-4503-8096-6/21/05}

\usepackage{adjustbox} 
\usepackage{balance}  
\usepackage{comment}
\usepackage{graphicx}
\usepackage{tabularx} 
\usepackage[caption=false, font=tiny, textfont=tiny]{subfig} 
\usepackage[inline]{enumitem} 
\usepackage{multirow}
\usepackage{csquotes}
\usepackage[nohyperlinks, nolist]{acronym}
\usepackage{todonotes}

\makeatletter
\def\url@leostyle{%
  \@ifundefined{selectfont}{\def\UrlFont{\sf}}{\def\UrlFont{\small\bf\ttfamily}}}
\makeatother
\def\pprw{8.5in}
\def\pprh{11in}

\newcommand{\peta}[1]{$\eta_p^2{=}#1$}

\newcommand{\fdf}[3]{$F_{#1,#2}{=}#3$}

\newcommand{\tOne}[2]{$t_{#1}{=}#2$}
\newcommand{\psig}[2][=]{$p{#1}#2$}

\newcommand{\cohensD}[1]{Cohen's $d{=}#1$}
\newcommand{\MeanSd}[2]{$M{=}#1$, $SD{=}#2$}

\begin{document}

\begin{acronym}
\acro{AR}{Augmented Reality}
\acro{BIP}{Break in Presence}
\acro{inVRQ}{in-VR Questionnaire}
\acro{HUD}{Head-up-Display}
\acro{HMD}{Head-Mounted Display}
\acro{MR}{Mixed Reality}
\acro{outVRQ}{out-VR questionnaire}
\acro{RQOne}{\textsc{RQ1}}
\acro{RQTwo}{\textsc{RQ2}}
\acro{RQThree}{\textsc{RQ3}}
\acro{VE}{Virtual Environment}
\acro{VR}{Virtual Reality}
\acro{WPM}{Words per Minute}
\acro{UI}{User Interface}
\acro{UX}{User Experience}
\end{acronym}

\title{Towards Low-burden Responses to Open Questions in VR}

\author{Dmitry Alexandrovsky}
\email{dimi@uni-bremen.de}
\affiliation{%
  \institution{DMLab, University of Bremen}
  \city{Bremen}
  \postcode{43017-6221}
  \country{Germany}
}

\author{Susanne Putze}
\email{sputze@uni-bremen.de}
\affiliation{%
  \institution{DMLab, University of Bremen}
  \city{Bremen}
  \country{Germany}
}

\author{Alexander Schülke}
\email{schuelk1@uni-bremen.de}
\affiliation{%
  \institution{University of Bremen}
  \city{Bremen}
  \country{Germany}
}

\author{Rainer Malaka}
\email{malaka@tzi.de}
\affiliation{%
  \institution{DMLab, University of Bremen}
  \city{Bremen}
  \country{Germany}
}

\begin{abstract}
Subjective self-reports in VR user studies is a burdening and often tedious task for the participants. To minimize the disruption with the ongoing experience VR research has started to administer the surveying directly inside the virtual environments. However, due to the tedious nature of text-entry in VR, most VR surveying tools focus on closed questions with predetermined responses, while open questions with free-text responses remain unexplored. This neglects a crucial part of UX research. To provide guidance on suitable self-reporting methods for open questions in VR user studies, this position paper presents a comparative study with three text-entry methods in VR and outlines future directions towards low-burden qualitative responding.
\end{abstract}

\ccsdesc[500]{Human-centered computing~HCI theory, concepts and models}
\ccsdesc[500]{Human-centered computing~Empirical studies in HCI}

\keywords{
    Virtual reality; self-reporting; design space; typing in VR.
}

\begin{teaserfigure}
    \centering
    \subfloat[Controller Drumming\label{fig:drumming}]{\includegraphics[width=0.3\textwidth]{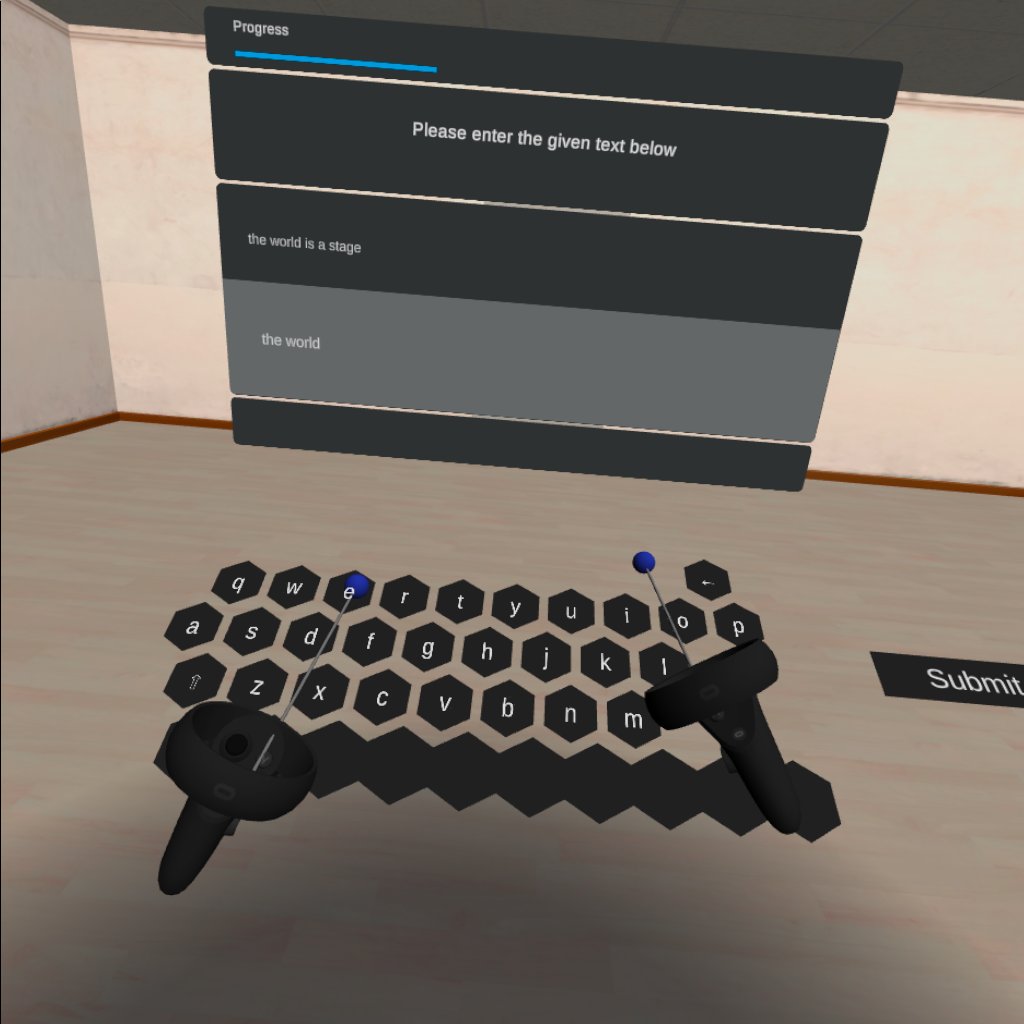}}
    \hspace{0.01\textwidth}
    \subfloat[Freehand\label{fig:freehand}]{\includegraphics[width=0.3\textwidth]{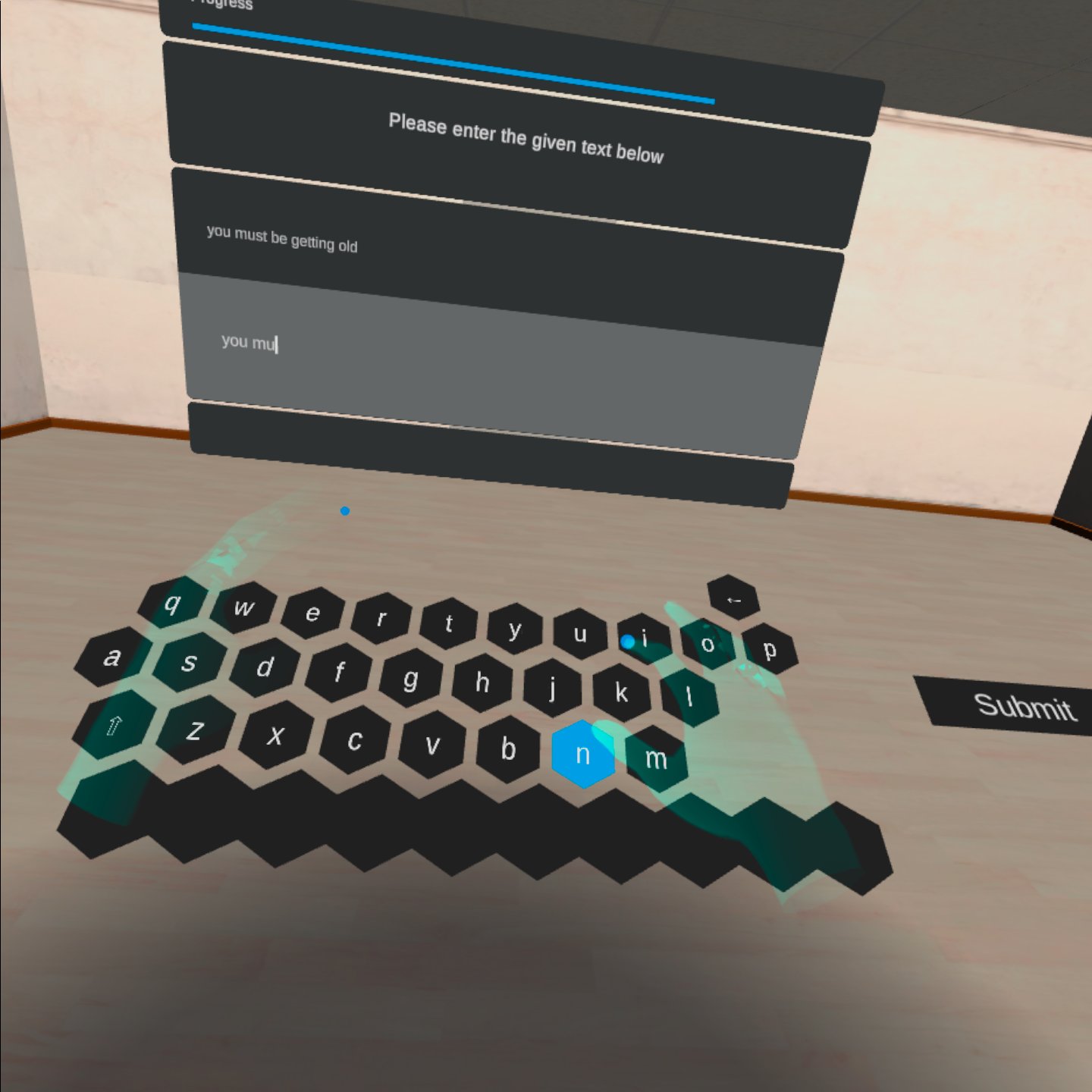}}
    \hspace{0.01\textwidth}
    \subfloat[Pinch\label{fig:pinch}]{\includegraphics[width=0.3\textwidth]{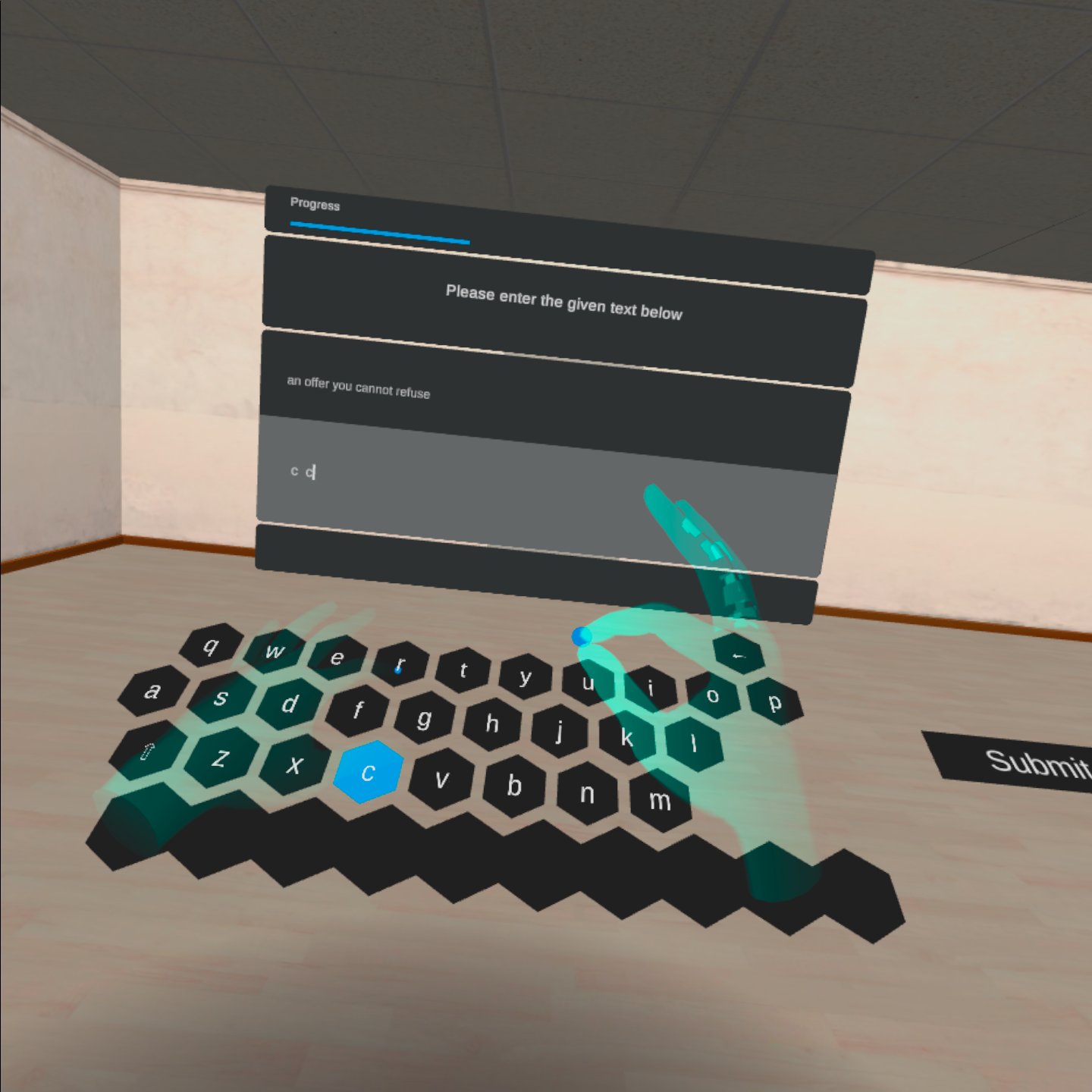}}
    \caption{The three input methods for open questions investigated in our user study.}
    \label{fig:text_input_methods}
\end{teaserfigure}

\maketitle


\input{body.tex}

\bibliography{VRQuestionnaires}

\bibliographystyle{ACM-Reference-Format}


\end{document}

%% file: body.tex
\section{Introduction and Related Work}

For \ac{VR} user studies, most setups rely on mid- or post-experience questionnaires outside \ac{VR}~\citep{skarbez2017, schwind2019}. Usually, after a tasks participants are required to leave \ac{VR} and fill out questionnaires on PC or paper. However, this change in realities breaks the study flow and produces a \ac{BIP}~\cite{slater2000, slater2003} which has been associated with disorientation, loss of control and negative emotions~\cite{knibbe2018,scherer2007, schwind2019,slater2003}. \citeauthor{putze2020} have shown that the switching between realities has a strong physiological effect which holds on for a significant period of time~\cite{putze2020}. 
Therefore, \acp{BIP} in \ac{VR} user studies might lead to uncontrolled biases, in particular right before the questionnaire responses. 
Moreover, particularly for unsupervised remote (i.e., online) \ac{VR} studies, the break of the study flow may cause the participants to drop off before finishing the experiment.
\ac{VR} research has started administering questionnaires inside the \ac{VE}; therefore, the participants do not need to leave the \ac{VE} to fill out questionnaires and can stay closer to the experience which improves the study flow along with the \ac{UX} of the the study~\cite{alexandrovsky2020,regal2019,schwind2019}.
However, the process of filling out questionnaires is interrupting and burdening, and in many cases it causes participants to terminate the study~\cite{moller2013, vanberkel2020}.
Although research has shown that \acp{inVRQ} minimize the \acp{BIP} in \ac{VR} user studies, they cannot avoid \acp{BIP} resulted by the question asking entirely~\cite{putze2020}.

\citeauthor{schwind2019}~\cite{schwind2019} contrasted the screen-based questionnaires against \ac{VR}-embedded questionnaires and found that with embedded assessment the subjective responses in \ac{VR} are more consistent. In contrast, others have shown that in-\ac{VR} questionnaires may lead to inconsistencies~\cite{graf2020}.
To counteract for such inconsistencies, \citeauthor{alexandrovsky2020}~\cite{alexandrovsky2020} presented important usability criteria for in-\ac{VR} questionnaires.
Other tools that allow administering questionnaires in \ac{VR} are the \ac{VR} Questionnaire Toolkit~\cite{feick2020}, \ac{VR}ate~\cite{regal2019}. 
Similarly,  MRAT~\cite{nebeling2020} is a toolkit for \ac{AR} studies.
These tools aim for a less-disruptive study flow and target problems of context-dependent forgetting~\cite{abernethy1940, godden1975} due to environment change~\cite{pohl2004} which may bias responses.
However, to date only closed questions (i.e. mostly Likert-scales) has been implemented and evaluated in \ac{VR}.
Nonetheless, qualitative research methods play a significant role in HCI as they often yield an explanation of quantitative outcomes and thus, provide for a deeper understanding of many phenomena~\cite{adams2008, blandford2016, lazar2017}. For instance, in post-experience semi-structured interviews or in surveys with open-ended questions, the participants could explain why the gave a good or a bad usability score when testing a system~\cite{}.
HCI shares a variety of qualitative research methods including, but not limited to semi-structured interviews, think-aloud protocols, semi-open questions~\cite{lazar2017}. All these methods have there specific use cases and differ in the workload the participants exhibit. For example, \citeauthor{vandenhaak2003}~\cite{vandenhaak2003} concurrent and retrospective think-aloud protocols. The authors showed that both research methods uncover usability issues similar with similar quality, but the two assessment methods differ in how the usability problems are detected and in their impact on the participants' performance. Therefore, \citeauthor{vandenhaak2003} argue that concurrent think-aloud methods are not suitable for complex and demanding tasks, and in such scenarios post-experience responses would provide more reliable results.

To cover a broad range of research methods in \ac{VR}, contemporary \ac{VR}-embedded questionnaire toolkits should support low-burdening qualitative responses along with quantitative questions.
For qualitative assessments, questionnaires often employ open-ended questions where the participants are asked to elaborate on their rating, to provide additional explanations, or to give further feedback.
For \ac{VR} user studies, open questions with free-text responses remain unexplored. 
Partially, this is due to the fact that, despite a growing body of work, typing in \ac{VR} is slower and more tedious than typing on a real physical keyboard~\cite{speicher2018}. \citeauthor{speicher2018}~\cite{speicher2018} developed a design space of text input methods in \ac{VR}. The authors compared six different text input techniques in \ac{VR}. Namely: head pointing, controller pointing, controller tapping, freehand and discrete \& continuous cursor. The study results show clearly that pointing with hand-held controllers exceeds the other text input methods. However, the authors conclude that none of the virtual typing methods is able to compete with physical keyboards.
Using physical keyboards in \ac{VR} has been shown effective. Yet, such mixed reality approaches come at great expense since the setups introduce additional sensors \citep{knierim2018, pham2019} and are prone to failure. This makes such setups unsuitable for remote or field studies. 
Nevertheless, virtual typing in \ac{VR} exists in several consumer products such as the Oculus Quest menu, many \ac{VR} social platforms, and \ac{VR} games. Therefore, despite its inferiority, virtual typing is frequently used by end-users. Nonetheless, while different typing methods in \ac{VR} show comparable performances, it remains unclear if study participants are willing to use them and how the typing paradigm affects the participants' compliance to give free-text answers.

To guide the design of low-burdening self-reporting, \citeauthor{yan2019} proposed a design-space of in-situ self-reporting~\cite{yan2019}. The design space builds on five requirements that aim do minimize the burden of self-reports:
\begin{enumerate*}
    \item \textit{Minimal Disruptiveness}: the self-reporting should not distract participants from their ongoing experience,
    \item \textit{Inclusiveness}: the self-reporting should prevent the participants embarrassment and maintain their privacy,
    \item \textit{Low-focus}: self-reporting should require low attention,
    \item \textit{Intuitiveness}: the interface should be self-explanatory and easy to use, and
    \item \textit{Expressivity} the self-reporting interface should support a variety of question types
\end{enumerate*}.
While this framework provides valuable insights for desktop and mobile user studies, it misses some opportunities and tradeoffs that emerge with reality-altering interfaces.
To provide an initial step towards the design of self-reporting for open questions and to lay a groundwork for a future design space of self-reporting in \ac{VR}, this work follows the research question \textit{what type of existing virtual keyboards is most suitable for free-text responses in \ac{VR} user studies}. 
This paper presents a comparative study of different typing methods in \ac{VR} and outlines a planned study, which should compare different response modalities that should contribute  the standardization of self-reporting methods in \ac{VR} user studies.

\section{Evaluating Suitable VR Text Input Methods}
\label{sec:text_input_methods}

Based on literature and existing end-user \ac{VR} applications we implemented three typing methods. Namely: \textit{Controller Drumming}, \textit{Freehand Typing}, and \textit{Pinch Typing}. Previous research compared these inputs with other existing methods and showed comparable performances and usability which makes them the best candidates for low-burdening text input for free-text responses in \ac{VR} user studies~\cite{dube2019, fashimpaur2020, speicher2018}.
To provide a sound comparability, all input methods were operating on the same virtual keyboard.
The participants could select from two keyboard layouts: QUERTY and QUERTZ.
The keyboard has a total area of $23{\times}60 cm$ with a hexagonal shaped keys with diameter of $8 cm$. This size determined by recommendations from previous work by \citeauthor{dudley2019}~\cite{dudley2019} and internal testing during the development.
Users can adjust the height and position of the keyboard.

\paragraph{Controller Drumming}
Our controller-based method follows \citeauthor{boletsis2019}'s~\cite{boletsis2019} implementation of drum-like text input (c.f., Fig.~\ref{fig:drumming}). This is a symmetric bimanual typing technique with \ac{VR} controllers. At each controller is a $14 cm$ long front-facing drum stick attached. The users performs keystrokes by hitting keys with the sticks.

\paragraph{Freehand}
The Freehand typing employs of the Oculus Quest's hand tracking (c.f., Fig.~\ref{fig:freehand}). Inside \ac{VR}, the users' hands are represented by high-detail meshes. Like controller drumming, this method is bimanually symmetric. At each index finger is a small sphere-shaped (r=$1.5 cm$) interaction area attached. The users perform the keystrokes by hitting keys with their index fingers. In line with findings by \citeauthor{dudley2019}~\cite{dudley2019} who demonstrated that index-only typing provides better performance, we implemented freehand typing with index finger only keystrokes.

\paragraph{Pinch}
This input method is inspired by PinchType~\cite{fashimpaur2020}.
Like freehand, this technique is build upon the hand tracking. However, this method is asymmetric  (c.f., Fig.~\ref{fig:pinch}). Like with the freehand method, the index finger of the dominant hand is used for key selection. However, in contrast to the other methods the key selection is not determined by collision of the hand and the key, but rather through proximity; i.e., the closed key to the index finger is highlighted. The actual keystroke is performed with a pinch gesture of the non-dominant hand.



\subsection{Study Design}

The goal of the study was to determine the usability of the three concurrent text input methods. Since the typing speed differs drastically between individuals we conducted the study within-subjects. The order of the conditions was randomized using Latin square.
We conducted the user study remotely. The participants downloaded the study app and participated at home using an Oculus Quest \ac{HMD}. The study app guided the participants through the study flow without any communication with an experimenter.
For each condition, the task was to write off sentences from the Enron corpus~\cite{vertanen2011} of standardized phrases with a length of $20{-}28$ characters. The target phrases were displayed on a world-anchored \ac{UI} in \ac{VR} using the respective input method.

As objective measures of performance we logged each key stroke along with its timestamp, from which we could later calculate the participants' accuracy and the \ac{WPM}.
We assessed workload on a raw NASA-TLX~\cite{hart1986} and usability on UMUX~\cite{finstad2010} using \acp{inVRQ} by \citeauthor{alexandrovsky2020}~\cite{alexandrovsky2020}.
Afterward, the participants left the \ac{VE} to answer a conclusive questionnaire and perform a typing test in order to determine their \ac{WPM} on a regular keyboard.
For each typing variant, the participants were briefed. Next, they wrote five sentences with the respective input method and rated workload as well as the usability of the typing method.
The total study duration was \MeanSd{21}{4.88} minutes.

\subsection{Outcomes}
\begin{figure}
    \centering
    \subfloat[UMUX\label{fig:umux}]{\includegraphics[width=\linewidth]{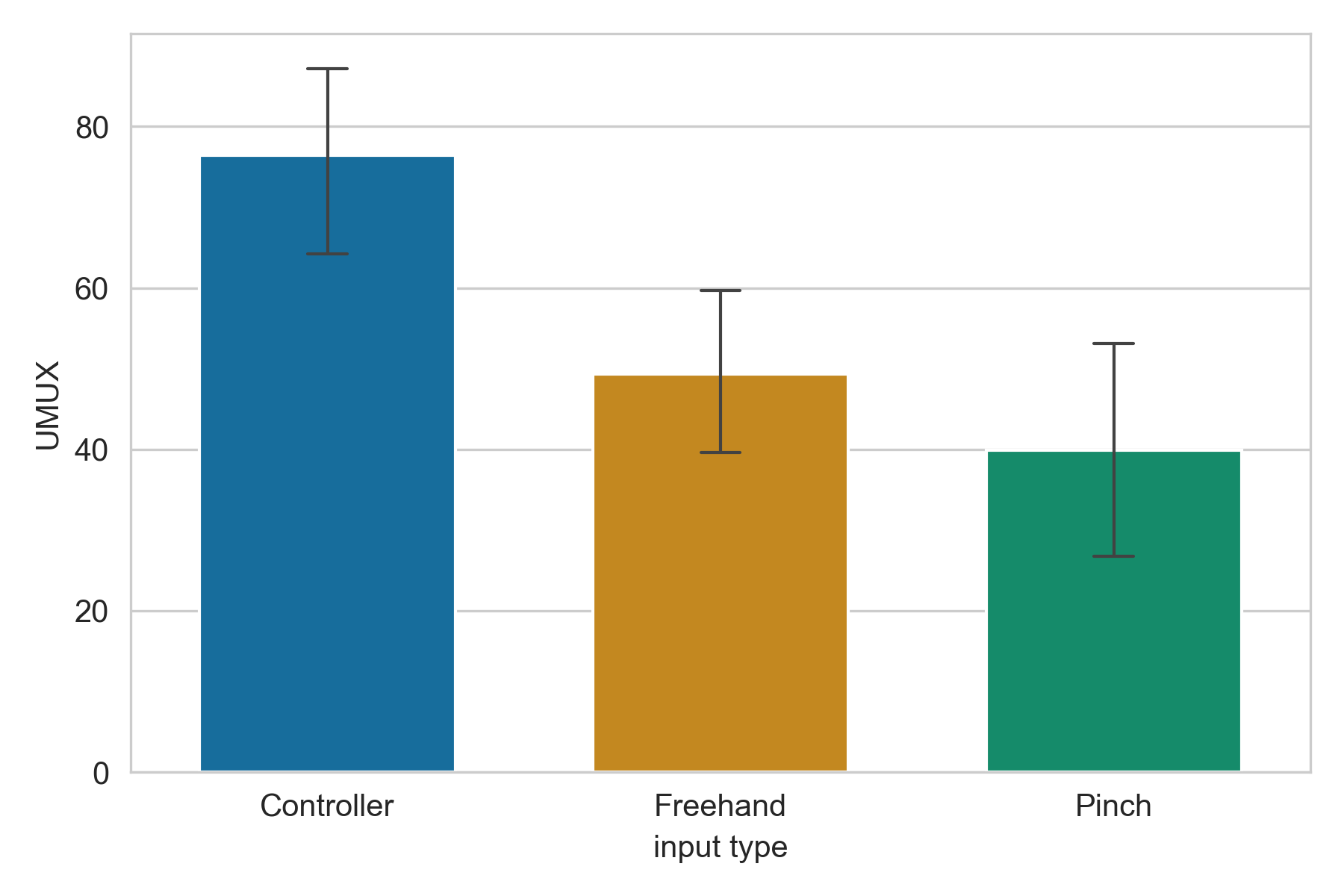}}
    
    \subfloat[NASA-TLX\label{fig:tlx}]{\includegraphics[width=\linewidth]{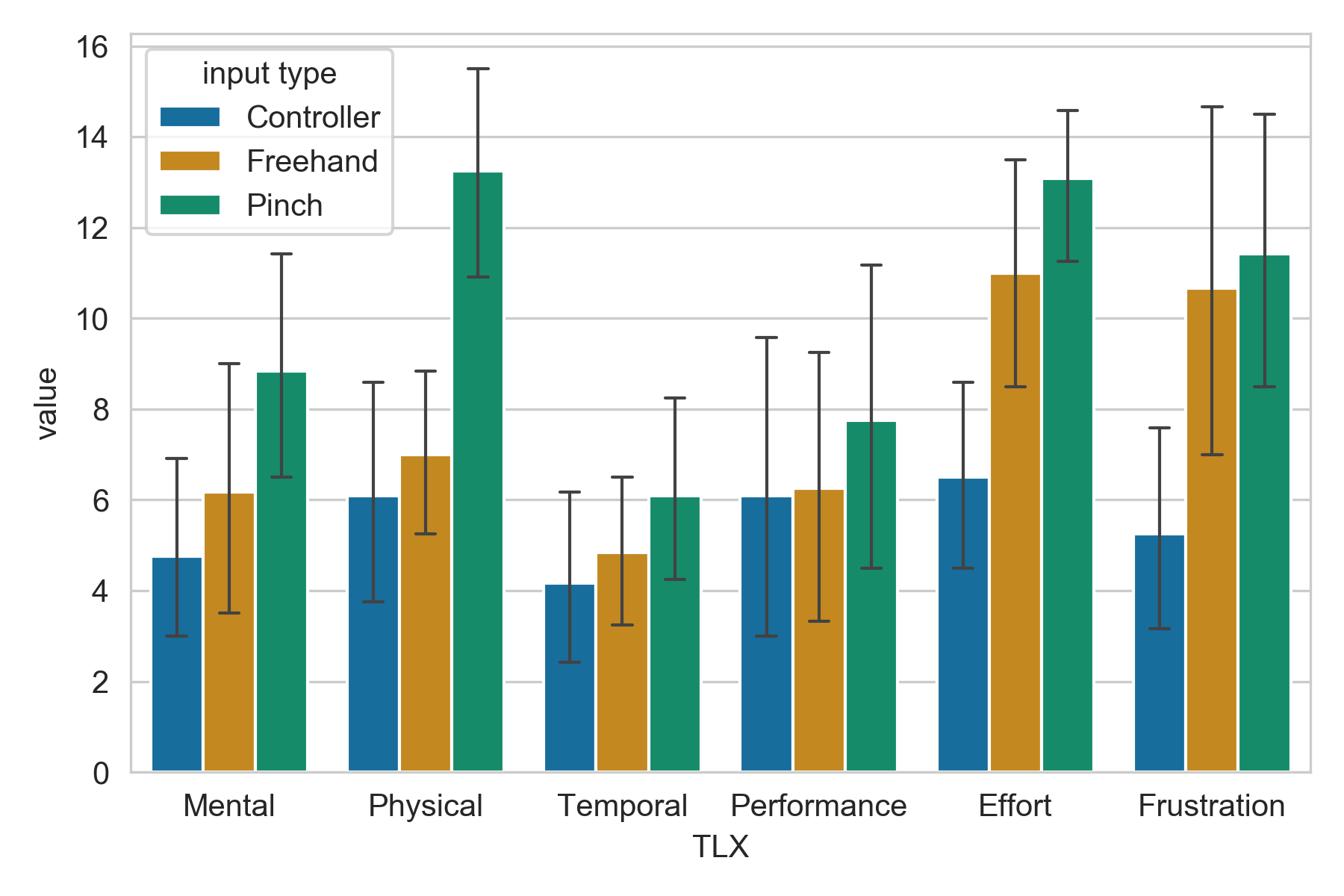}}
    
    \subfloat[Net WPM\label{fig:wpm}]{\includegraphics[width=\linewidth]{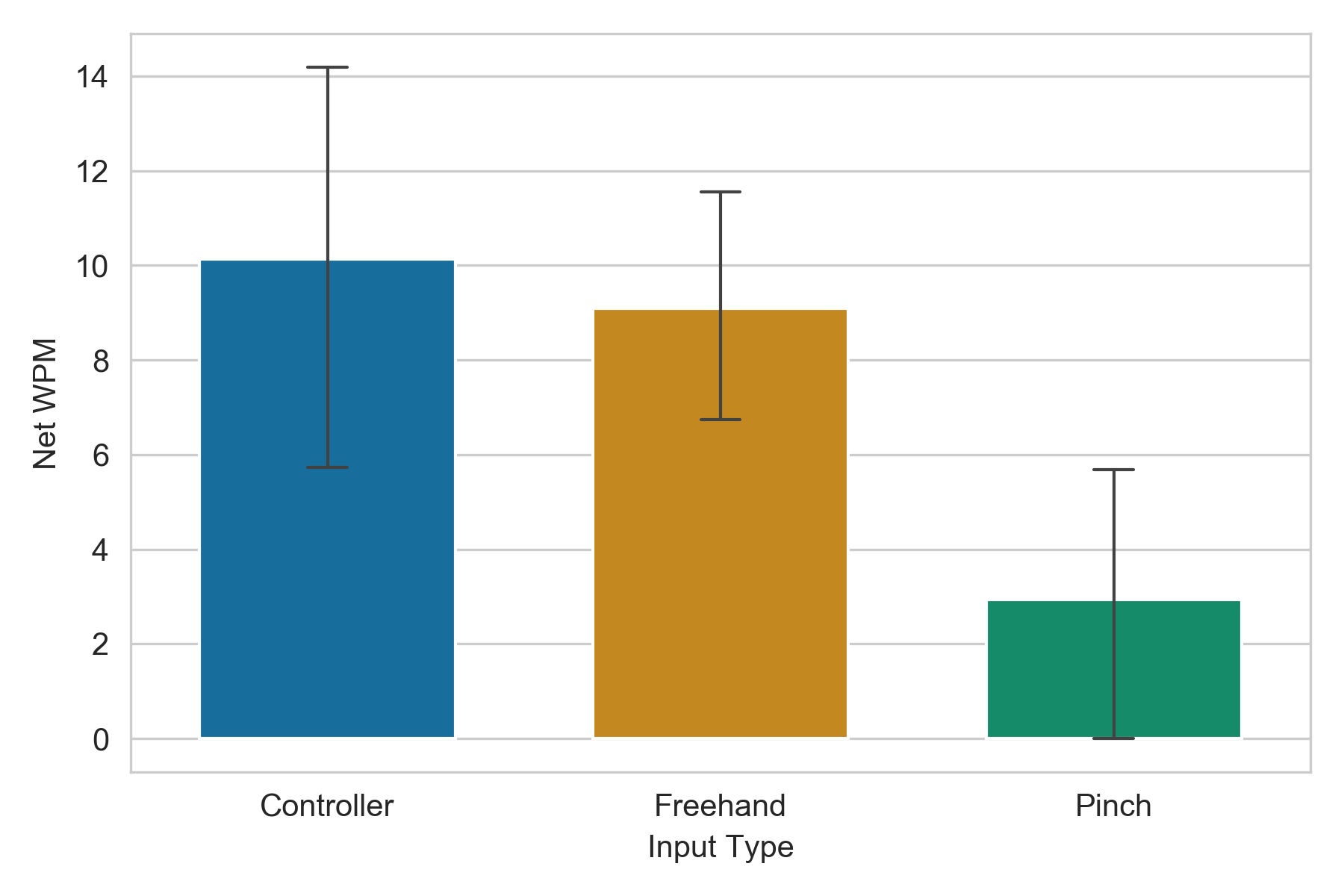}}
    
    \caption{User study results.}
    \label{fig:pre-study_plots}
\end{figure}

$12$ participants (self-reported: $8$ male, $4$ female, age: $M{=}27.08$ ($SD{=}1.38$)) volunteered for our study.
The typing speed on a regular keyboard out-\ac{VR} was \MeanSd{63.58}{22.13}.
To determine differences between the typing variants, we conducted RM-ANOVAs with the typing method as within-subjects factor on all measures.
The usability on the UMUX (Fig.~\ref{fig:umux}) differed significantly, between the input methods (\fdf{2}{22}{14.03}, \psig[<]{0.01}, \peta{0.56}). Post-hoc comparisons revealed that drumming receives significantly higher scores over freehand (\tOne{11}{4.29}, \psig[<]{0.01}, \cohensD{1.39}) and over pinch (\tOne{11}{5.29}, \psig[<]{0.01}, \cohensD{1.59}). 
The workload on TLX (Fig.~\ref{fig:tlx}) differed significantly for mental demand (\fdf{2}{22}{7.95}, \psig[<]{0.01}, \peta{0.42}) physical demand (\fdf{2}{22}{28.90}, \psig[<]{0.01}, \peta{0.72}), effort (\fdf{2}{22}{20.07}, \psig[<]{0.01}, \peta{0.65}) and frustration (\fdf{2}{22}{9.31}, \psig[<]{0.01}, \peta{0.46}), showing that generally the controller based input method promoted the lowest workload.
For the performance analysis, two participants were excluded due to corrupt or missing data. The Net \ac{WPM} (Fig.~\ref{fig:wpm}), i.e.,  raw \ac{WPM} reduced by the uncorrected errors, differed significantly between the conditions (\fdf{2}{18}{20.42}, \psig[<]{0.01}, \peta{0.69}) showing that the drumming was significantly faster compared to pinch (\tOne{49}{2.79}, \psig{0.02}, \cohensD{0.54}) and that freehand was significantly faster than pinch (\tOne{49}{3.06}, \psig{0.01}, \cohensD{0.61}).
These corroborated results show clearly that controller based typing is the most suitable text-entry method. Furthermore, these results are in line with related literature confirming that typing in \ac{VR} by a magnitude slower than out-\ac{VR}.

\section{Towards Low-Burdening Open Questions Responses in VR}
\label{sec:discussion}
The outcomes from the comparison study of different text-entry methods in \ac{VR} underline that typing in \ac{VR} is a difficult and burdening task. Most likely the burdening typing methods would disengage the participants from an ongoing user study leading to missing data, incorrect, or incomplete responses to open questions. To address these issues and to provide a clearer picture on the participants' response behavior we plan a user study which compares the compliance of participants of responding to open questions using different input methods. The study is planned as between-subjects design with the conditions
\begin{enumerate*}
    \item \ac{VR} drumming as the best suitable text-entry method in \ac{VR},
    \item voice recordings of the responses in \ac{VR} and as an low-burdening and intuitive interaction,
    \item a \enquote{traditional} text-entry method using \acp{outVRQ}
\end{enumerate*}.

To avoid biases, the participants are decepted and tasked to play different variants of a simple shooter to provide qualitative responses about what they liked or disliked about the game. As a primary measure of compliance, we aim to operationalize the response length.

The to be explored modalities cover a wide range of tradeoffs for the study design which require careful consideration.
The outcomes of the planned study should give insights about the participants' compliance on \citeauthor{yan2019}'s dimensions of \textit{Disruptiveness}, \textit{Inclusiveness}, and \textit{Intuitiveness}. Furthermore, the study should contribute to the foundation of a design space of low-burden self-reporting in \ac{VR} which aims to classify the benefits and drawbacks of different design decisions regarding self-reporting in \ac{VR} user studies and to provide a groundwork for standardized methods for \ac{VR} research.